\pdfoutput=1
\documentclass[preprint,3p,twocolumn]{elsarticle}
\usepackage{graphicx}
\usepackage{pdfpages}
\usepackage{epstopdf}
\usepackage{hyperref}
\hypersetup{colorlinks = true}
\newcommand{\figwidth}{3.3in}

\journal{Nuclear Instruments and Methods A}
\begin{document}
\begin{frontmatter}
\title{Technique for Surface Background Rejection in Liquid Argon Dark Matter Detectors using Layered Wavelength-Shifting and Scintillating Thin Films }
\newcommand{\carleton}{Department of Physics, Carleton University, Ottawa, Ontario, Canada K1S 5B6}
\newcommand{\astrocent}{AstroCeNT, Nicolaus Copernicus Astronomical Center, Polish Academy of Sciences, Rektorska 4, 00-614 Warsaw, Poland}
\author[CU]{M.\,G.~Boulay} 
\author[AC,CU]{M.~Ku{\'z}niak}
\address[CU]{\carleton}
\address[AC]{\astrocent}

\begin{abstract}
A technique using layered wavelength shifting, scintillating and non-scintillating films is presented to achieve discrimination of surface $\alpha$ events from low-energy nuclear recoils in liquid argon detectors.  A discrimination power greater than $10^{8}$, similar to the discrimination possible for electronic recoils in argon, can be achieved by adding a 50~$\mu$m layer of scintillator with a suitably slow decay time, approximately 300~ns or greater, to a wavelength-shifter coated surface.  The technique would allow suppression of surface $\alpha$ events in a very large next-generation argon dark matter experiment (with hundreds of square meters of surface area) without the requirement for position reconstruction, thus allowing utilization of more of the instrumented mass in the dark matter search.  The technique could also be used to suppress surface backgrounds in compact argon detectors of low-energy nuclear recoils, for example in measurements of coherent neutrino-nucleus scattering or for sensitive measurements of neutron fluxes.
\end{abstract}
\begin{keyword}
dark matter \sep argon \sep scintillation \sep pulse-shape discrimination \sep alpha backgrounds 
\PACS 95.35.+d \sep 95.85.Ry \sep 95.55.Vj \sep 29.40.Mc

\end{keyword}
\end{frontmatter}

\date{\today}

\section{Introduction}
\label{section1}
It is well established that most of the matter present in our universe is non-baryonic dark matter.  Direct searches for the interaction of dark matter particles with large target masses of noble liquids search for low-energy nuclear recoils, on the order of 10s of keV or lower.
Low-energy nuclear recoils or other low-energy particles are detected in noble liquids by observing scintillation light or charge, or both, released in the interaction ~\cite{first_argon_psd,deap1_psd,ds50,XENON}.  
A critical feature in all of these experiments is the control or identification of radioactive backgrounds that could otherwise mimic a low-energy nuclear recoil.

Dominant detector backgrounds are typically from electromagnetic interactions (from $\beta$'s or $\gamma$'s).  It has been demonstrated that electromagnetic backgrounds can be effectively suppressed in liquid argon using pulse-shape discrimination (PSD) based on differences of the time evolution of the scintillation pulse between electronic and nuclear recoils~\cite{first_argon_psd, deap1_psd}.  Assuming that backgrounds from sources external to the detector are mitigated with PSD or shielded, which is the main concept of the DEAP-3600 detector~\cite{deap_det}, another important source of background to low-energy nuclear recoil events is from $\alpha$-decays on or near detector surfaces.  Most naturally-occurring $\alpha$-emitters have energies of several MeV, well above the region of interest for low-energy nuclear recoils in  dark matter searches.  $\alpha$-decays occurring in the sensitive volume of noble liquid detectors are thus easily identified by their large energy deposition.  Near detector surfaces, however, an $\alpha$-decay can lead to a recoiling nucleus being ejected into the sensitive detector region, while the $\alpha$-particle itself travels toward the detector wall and is either not observed, or observed with a vastly lower apparent energy, possibly low enough to be confused with a signal.  Even if no recoil nucleus is ejected into the sensitive target region, an $\alpha$-particle which loses most of its energy in a non-active material and in so doing deposits only a small amount of energy in the sensitive detector region  could be mis-identified as a low-energy recoil.  Sensitive dark matter searches typically reject events from detector surfaces, including those borne by $\alpha$-decays, by reconstructing event positions and defining  an inner fiducial volume that is free of or lower in background events.

\begin{figure}
\begin{center}
\includegraphics[trim={1.3cm 0 0.5cm 0},clip,width=\columnwidth]{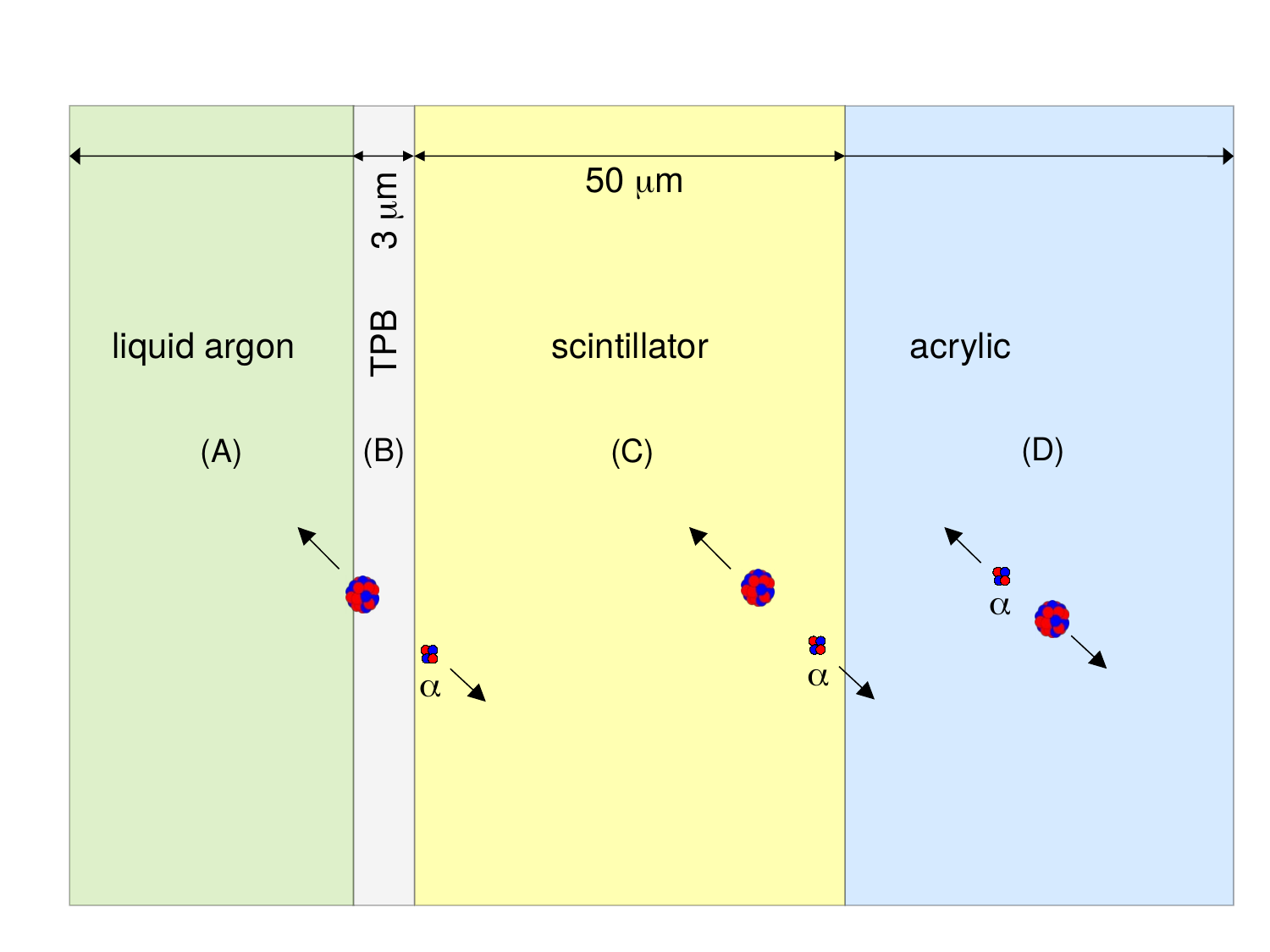}
\caption{\label{concept}Proposed configuration of multi-layer thin films at a detector surface.  $\alpha$-decays from regions A or B will always deposit all of their energy in an active layer, and can be discriminated from low-energy nuclear recoils based on the high observed energy.  Decays from region C will either deposit all of their energy in an active region (A, B or C) and be discriminated based on high deposited energy, or will deposit only a fraction of their energy in region C, in which case they can be discriminated against based on the long decay time of the scintillator.  Similarly, events originating in the acrylic can be discriminated based on the long scintillator decay time.}
\end{center}
\end{figure}

We present here an example design and simulation results demonstrating that a multi-layer arrangement of liquid argon, a thin film of tetraphenyl butadiene (TPB) wavelength shifter, a thin scintillator film with a sufficiently slow decay time constant, and an inactive but optically transparent region can be used to effectively identify this source of background events in liquid argon without requiring position reconstruction and requires only photon timing information.  The layout is shown in Fig.~\ref{concept}.  Nuclear recoils in the argon (region A) generate scintillation light, which is comprised of mostly prompt vacuum ultraviolet (VUV) photons~\cite{hitachi}.  These photons are wavelength shifted by a thin (3 micron) TPB layer (region B), which converts the prompt VUV scintillation pulse into a prompt pulse of visible photons.   The remaining layers, as well as the liquid argon, are transparent to the visible photons, which are ultimately transported through the scintillator and acrylic (regions C and D) to a photodetector (on the right of the acrylic in Fig.~\ref{concept}, not shown).  This mechanism of converting and detecting the prompt scintillation pulse is the same as used in most liquid argon detectors~\cite{ds50,deap_physics2,deap_physics1,cenns} and allows powerful discrimination of nuclear recoil events from slower pulses generated by electronic recoils in argon.   The concept presented here includes an additional scintillating layer, which is selected to have a slow decay time, and which generates visible scintillation photons from $\alpha$-particle interactions. The layer selected here is $50~\mu$m thick, thus able to stop all $\alpha$-particles from the naturally occurring uranium and thorium decay chains, entering from either side.  Photons generated in the scintillating layer will then also be ultimately observed by the photodetectors. The slow decay time of the scintillating layer will guarantee that even the problematic surface $\alpha$-decays with low observed energy will have very different pulse shapes than nuclear recoil events inside of the fiducial mass, and will be efficiently mitigated with PSD.  The acrylic layer shown is non-scintillating, and will simply absorb deposited $\alpha$-energy without any re-emission.  While acrylic used in this study is a common material choice for liquid argon containment in dark matter detectors, any other inactive and transparent material would yield similar results.  The overall assembly is similar to a ``phoswich'' or ``phosphor sandwich'' scintillator assembly used for $\alpha/\beta(\gamma)$ pulse-shape discrimination~\cite{phoswich1, phoswich2}, except in this case the layers are thin films, and the TPB layer is used to shift the wavelength of VUV photons.

Ionizing radiation in scintillators typically leads to the formation of excited singlet or triplet states.  These states will have characteristic lifetimes ($\tau_{singlet}$ and $\tau_{triplet}$)
and are produced with different amplitudes ($I_{singlet}$ and $I_{triplet}$), depending on the scintillator and the ionizing radiation.   In this study we have parameterized each scintillator with up to four components, each with a given decay lifetime and intensity, shown in Table~\ref{scint_pars}. Scintillation properties for layers A, B and D - liquid argon, TPB and acrylic (which does not scintillate) have all been studied extensively~\cite{deap1_psd,hitachi,tpb_psd_princeton,veloce_tpb,pollmann_tpb}.   Layer C, labeled as ``scintillator'' in the diagram is parameterized with a single time constant.  This can be considered an effective time constant for the scintillation pulse, relevant in the chosen data acquisition window, in the case of a scintillator with more than one significant slow emission time, and is useful in studying the effect of this time constant on the $\alpha$ rejection power.

\begin{table}
\caption{\label{scint_pars}Scintillation parameters for $\alpha$-particles used in this study.  Scintillation is parameterized with up to four components, with $I_{1}+I_{2}+I_{3}+I_{4}=1$.  The number of generated photons per unit energy is the product of the Yield and the Quenching.  The scintillator is parameterized with a single time constant.  Acrylic is transparent and does not scintillate. Literature values for TPB~\cite{veloce_tpb,pollmann_tpb} and argon~\cite{Doke_2002} are used.}
\begin{tabular}{lllll} 
Parameter                 & Region & Region & Region \\
                 & A (argon)       & B (TPB)      & C (scint.) \\ \hline
Yield (ph/keV)       & 40    & 0.882  & 18  \\
Quenching ($\alpha$) & 0.72 & 1  & 0.15  \\
$I_{1}$ &0.7 & 0.24 & 1   \\
$\tau_{1}$  & 6 ns        & 10.5 ns  & $\tau_{s}$            \\
$I_{2}$& 0.3 &0.089 &    \\
$\tau_{2}$        & 1500 ns                      & 800 ns           &    \\
$I_{3}$ &  & 0.233 &  \\
$\tau_{3}$   &                       & 6.5 $\mu$s          &  \\
$I_{4}$ &  &0.44 & \\
$\tau_{4}$         & & 39.9 $\mu$s                      &   \\
\end{tabular}
\end{table}

\section{Details of the simulation}
\label{details}
To evaluate background rejection, a fast Monte-Carlo simulation was implemented.  $\alpha$-decays of $^{210}$Po, a dominant source of problematic backgrounds, were simulated, generating $5.3$~MeV $\alpha$-particles and the recoiling $^{206}$Pb nuclei.  The simulation uses the stopping power in argon and in the detector wall, as calculated by SRIM~\cite{srim}.  Given the generality of this study and unspecified long time-constant scintillator material, for simplicity the stopping power for acrylic substrate was also used for TPB and for the slow scintillator. Since they would  be of the same order, this approximation has on impact on the conclusions of this work. Ions are tracked through all materials, a parameterization for straggling based on the SRIM calculations is included, and the energy deposited in each material is calculated.  Observed time distributions of photons are simulated using the deposited energies and the scintillator parameters from Table~\ref{scint_pars}.  The detector parameters used in the simulation are shown in Table~\ref{detsim}, which include the photon detection efficiency and the total noise rate, based on a large-area detector (200~m$^{2}$ surface area) containing typical photomultiplier tubes. Smaller detectors would have a better discrimination than calculated here, since the overall noise rates would be lower due to the lower number of photomultiplier tubes.  The simulation samples a Poisson distribution for the detected photons, assuming that none of the photons or noise hits are correlated.  This assumption has been shown to describe scintillation induced by electronic and nuclear recoils in argon very well~\cite{deap1_psd}.  We use here the same F$_{prompt}$ variable defined in \cite{first_argon_psd} as the PSD parameter, defined as the ratio of observed photons in the first 150~ns of the pulse to the full pulse, in this case defined as 16~$\mu$s.  The relatively wide 150~ns prompt window makes the PSD parameter insensitive to details of the detector trigger, and to reflections and scattering within the detector, for detectors up to several meters in size, so that the results presented here should be applicable to very large argon detectors.
\begin{table}
\caption{\label{detsim}Parameters used in the simulation.  Total photon detection efficiency and noise rates are typical values for a high photo-coverage of conventional photomultipliers, assuming a total surface area of 200 m$^{2}$.}
\begin{tabular}{ll} \hline
Parameter                 & Value \\ \hline
photon detection efficiency  & 20 \%   \\
dark noise hits (per 16 $\mu$s event) & 20   \\
LE NR quenching, $L_{eff}$  & 0.2\footnotemark \\
TPB WLS efficiency & 1 \\
\end{tabular}
\end{table}

\begin{figure}
\begin{center}
\includegraphics[width=\figwidth]{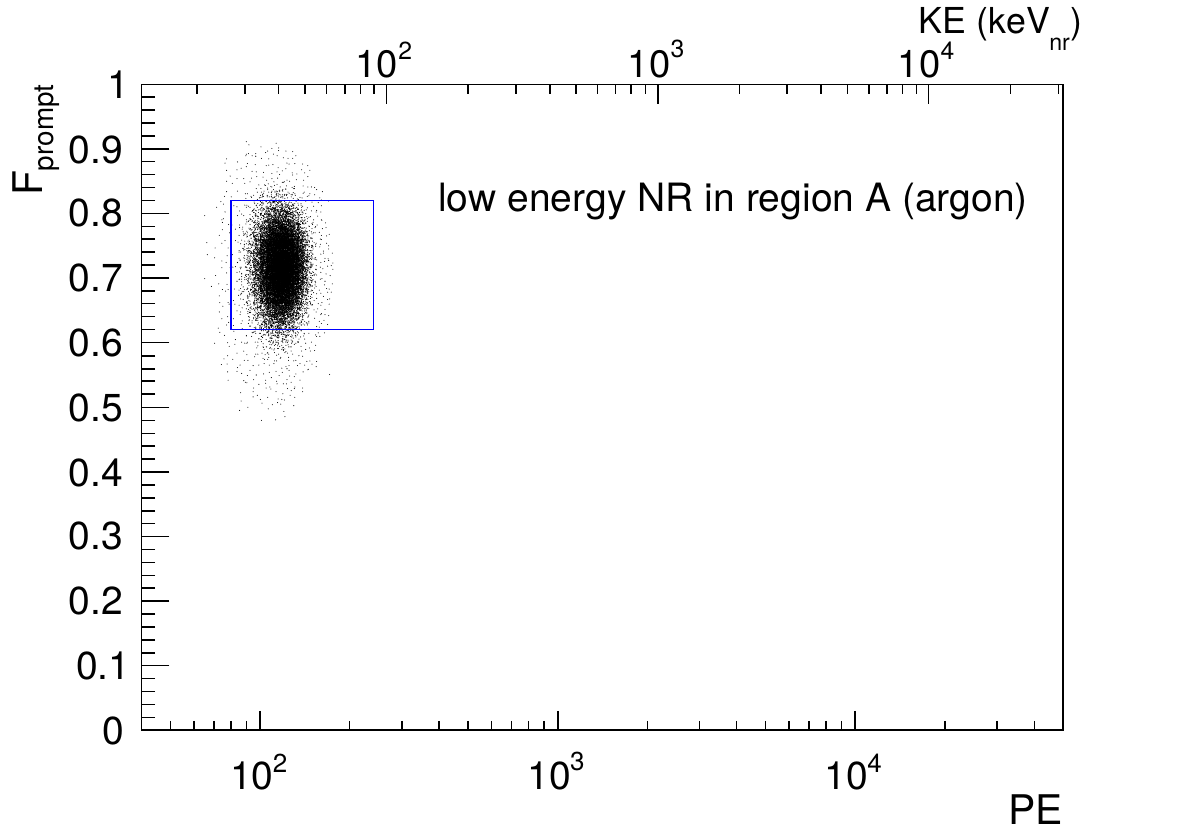}
\caption{\label{scint_nr}Low energy nuclear recoils in argon. The 99\% acceptance region defined by the box is used to evaluate background discrimination.}
\end{center}
\end{figure}

\begin{figure*}
\includegraphics[width=\figwidth]{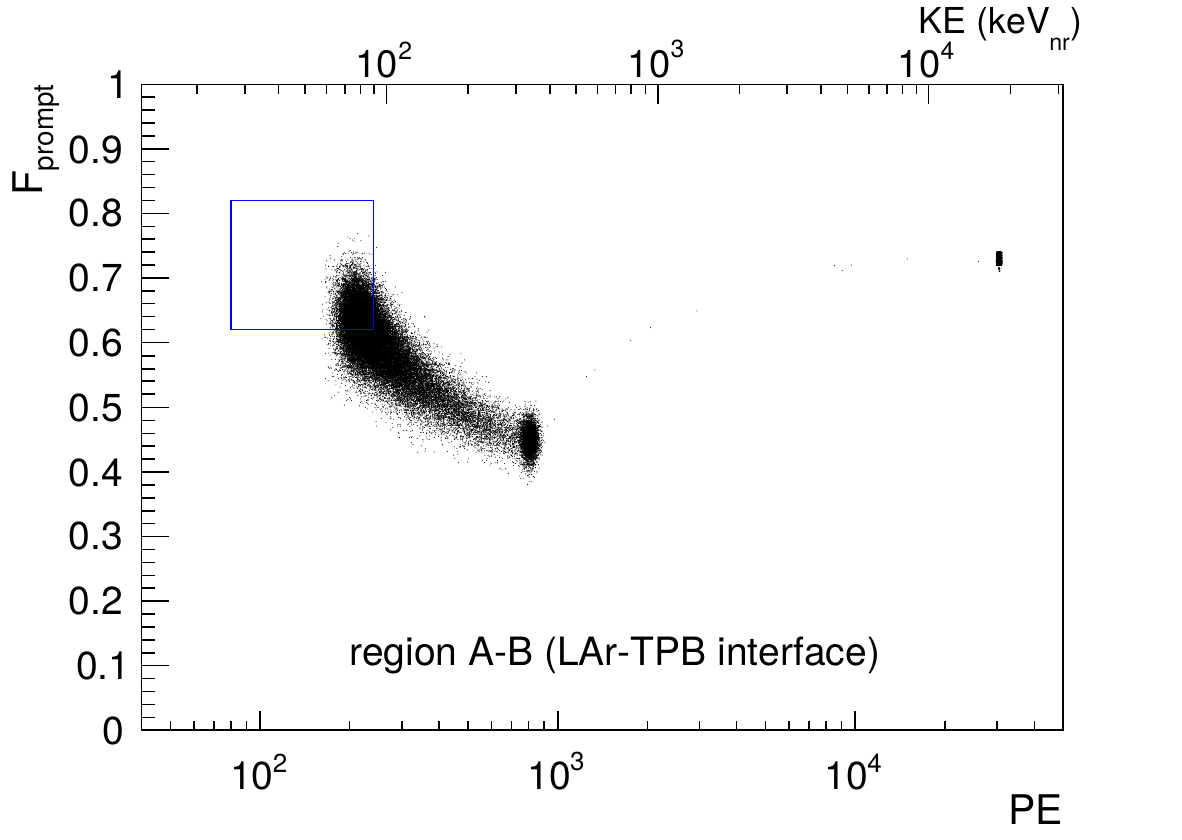}
\includegraphics[width=\figwidth]{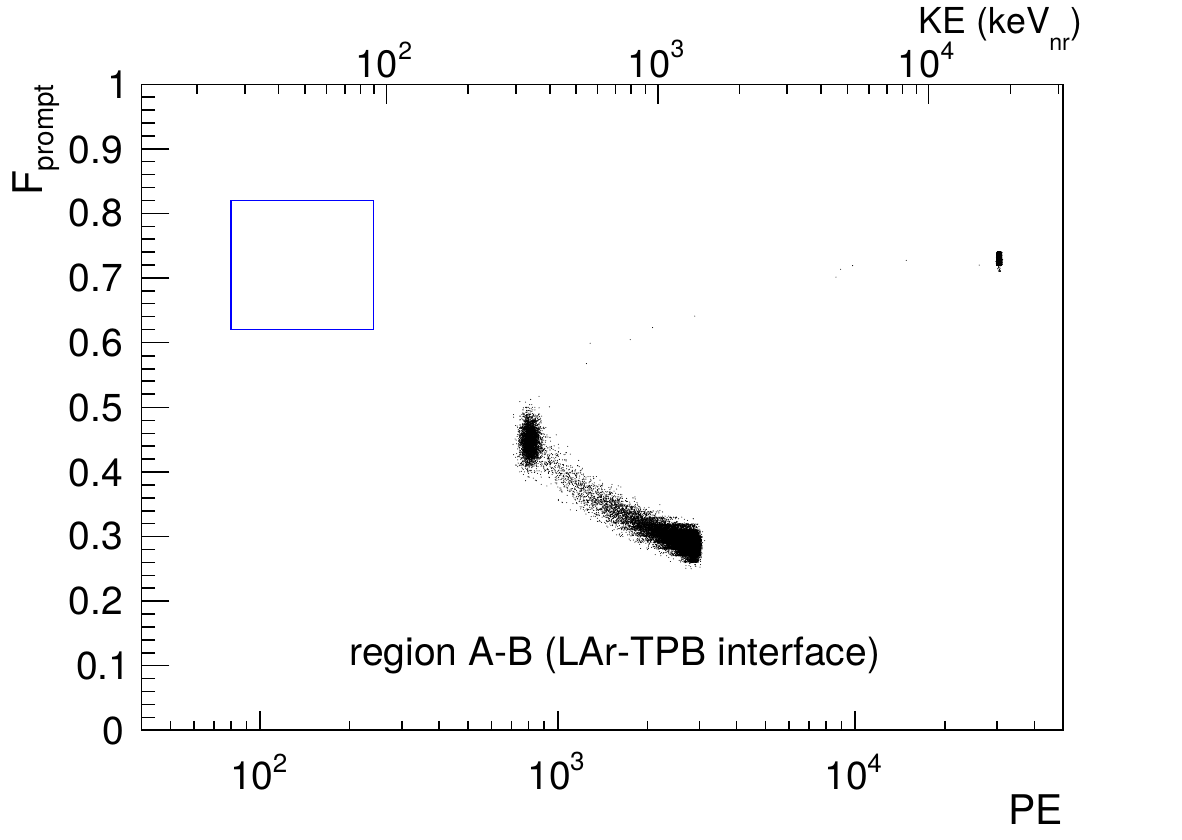}
\caption{\label{scint_tpb_surf} Simulation with events generated on the TPB surface: (left) without scintillating layer C.  In this case events can populate the region of interest, since nuclear recoils scintillating in liquid argon are prompt, and lead to a small energy deposition, with most of the $\alpha$ energy lost in the acrylic. (right) with scintillating layer C.  In this case, the $\alpha$-particle always loses all of its energy in an active material, and so the observed energy will be above the region of interest. In both cases the cluster of events at high PE and high F$_{prompt}$ corresponds to $\alpha$-particles depositing all of their energy in liquid argon.}
\end{figure*}

\begin{figure}
\includegraphics[width=\figwidth]{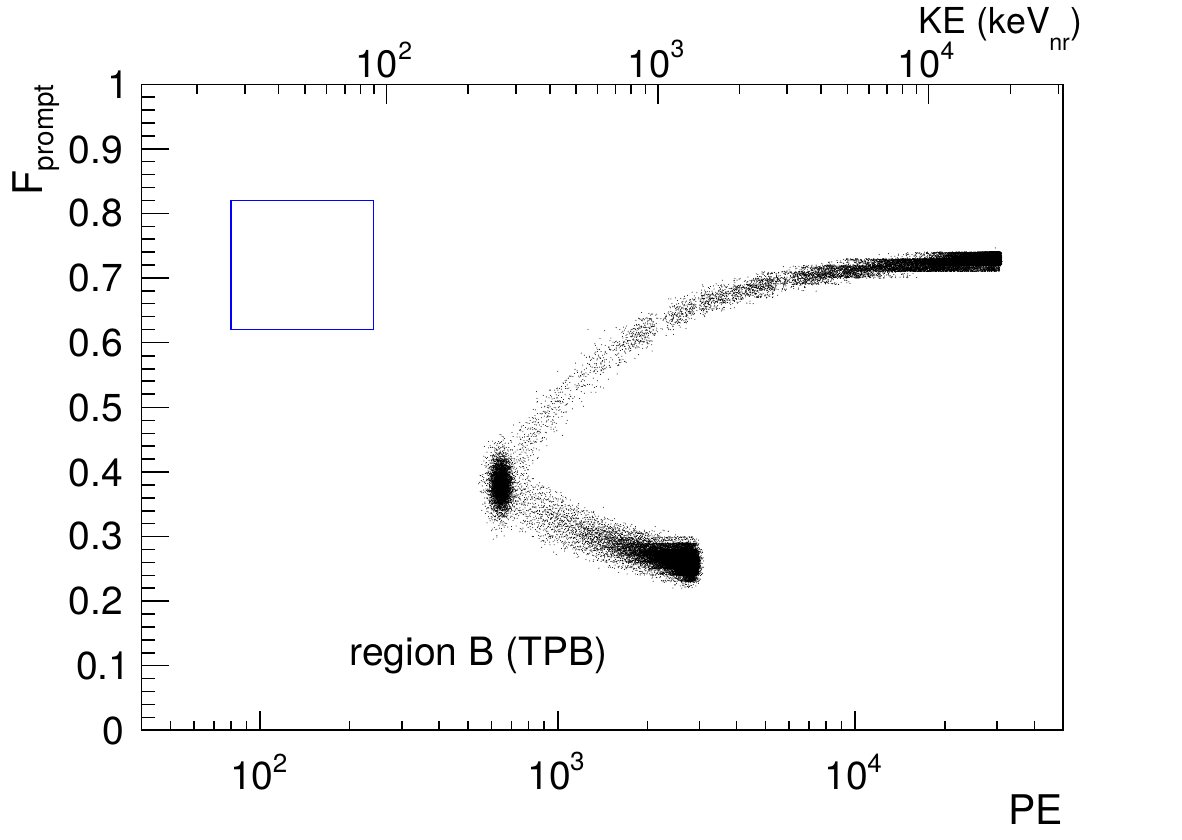}
\caption{\label{scint_tpb}Simulation of events generated in the TPB layer. The cluster of events at low PE contains decays with nuclear recoils depositing their entire energy in liquid argon and $\alpha$-particles traversing the TPB at close to normal incidence. In the low-F$_{prompt}$ band, increasing fraction and eventually all of $\alpha$-energy is deposited in TPB, as the track length increases with changing incidence angle.  The high-F$_{prompt}$ band contains events with increasing fraction of $\alpha$-energy deposited in liquid argon, which results in prompt scintillation. All events deposit energy in an active region, and so are above the low-energy region of interest.}
\end{figure}

\begin{figure*}
\includegraphics[width=\figwidth]{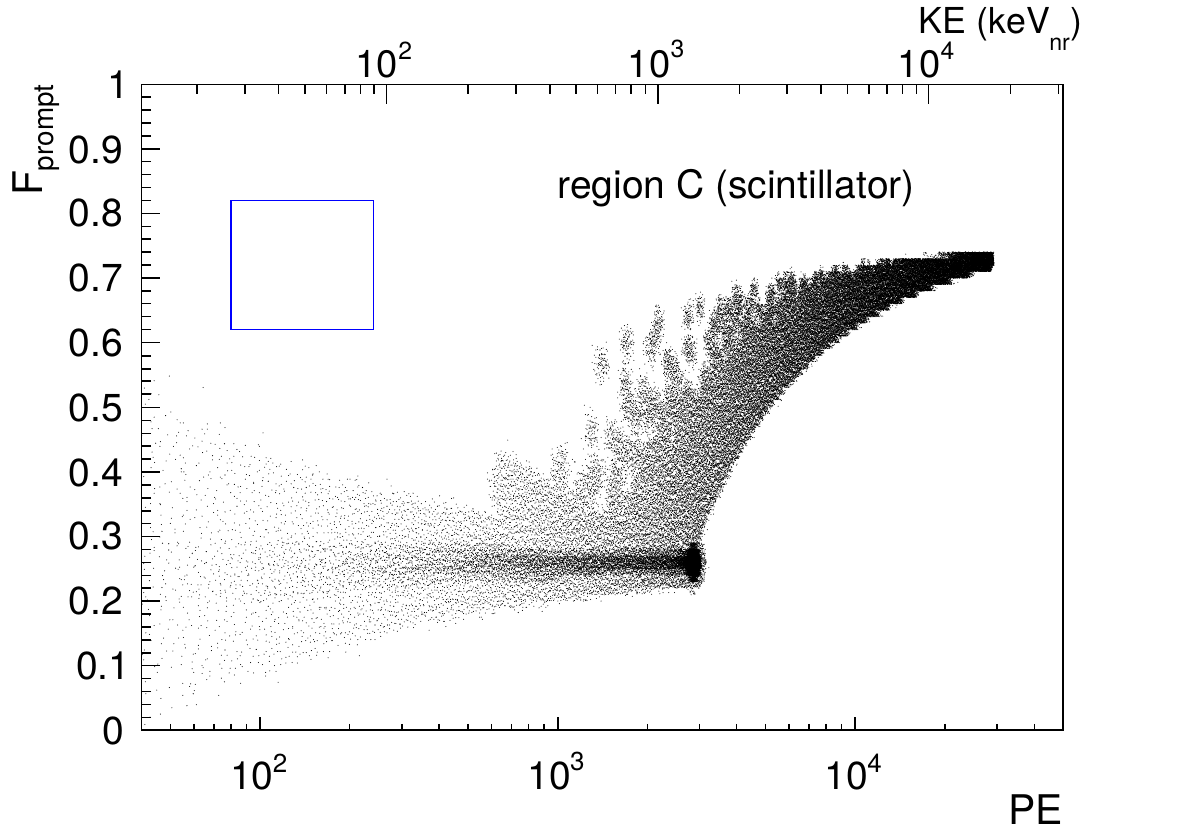}
\includegraphics[width=\figwidth]{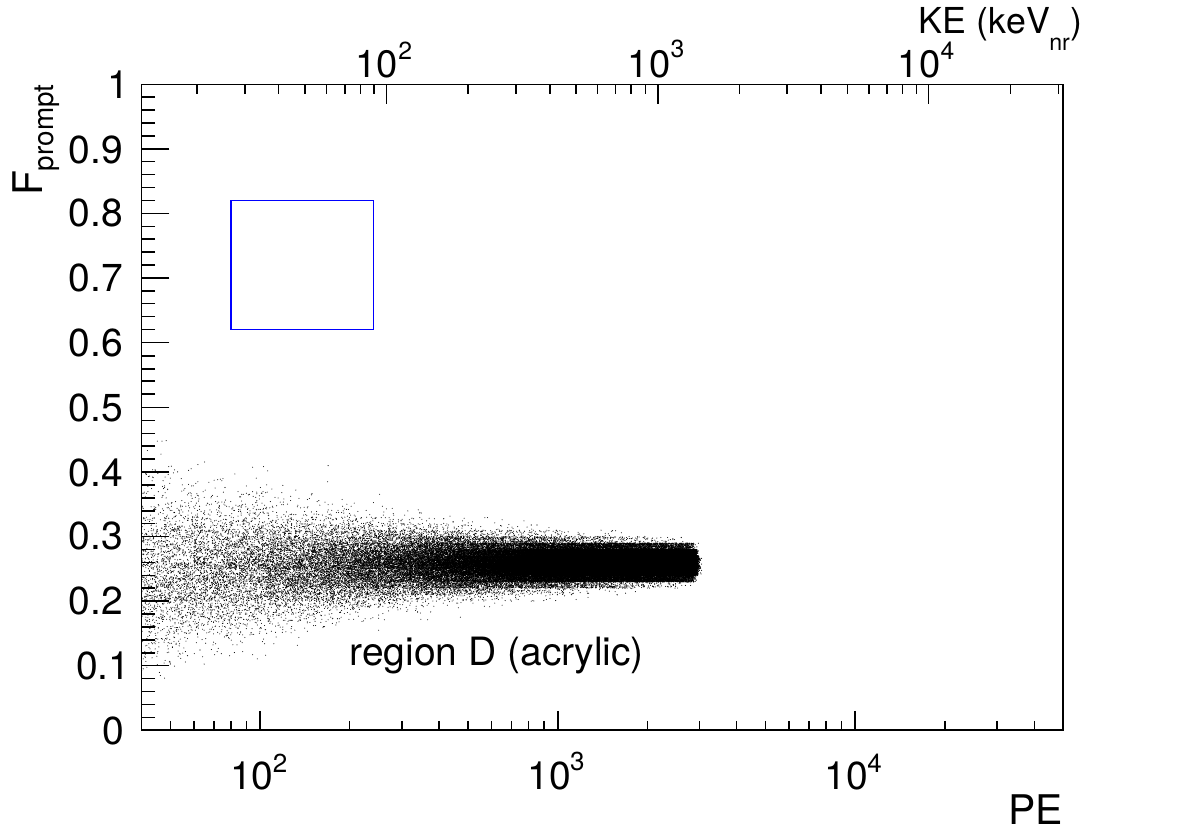}
\caption{\label{scint_scint}Simulated events in the scintillator layer (left) and in the acrylic (right).  In both of these cases the $\alpha$ can lose most of its energy in acrylic, and so the slow decay constant of the scintillator, leading to low F$_{prompt}$ values, is needed to discriminate events based on PSD.}
\end{figure*}

\begin{figure*}
\includegraphics[width=\figwidth]{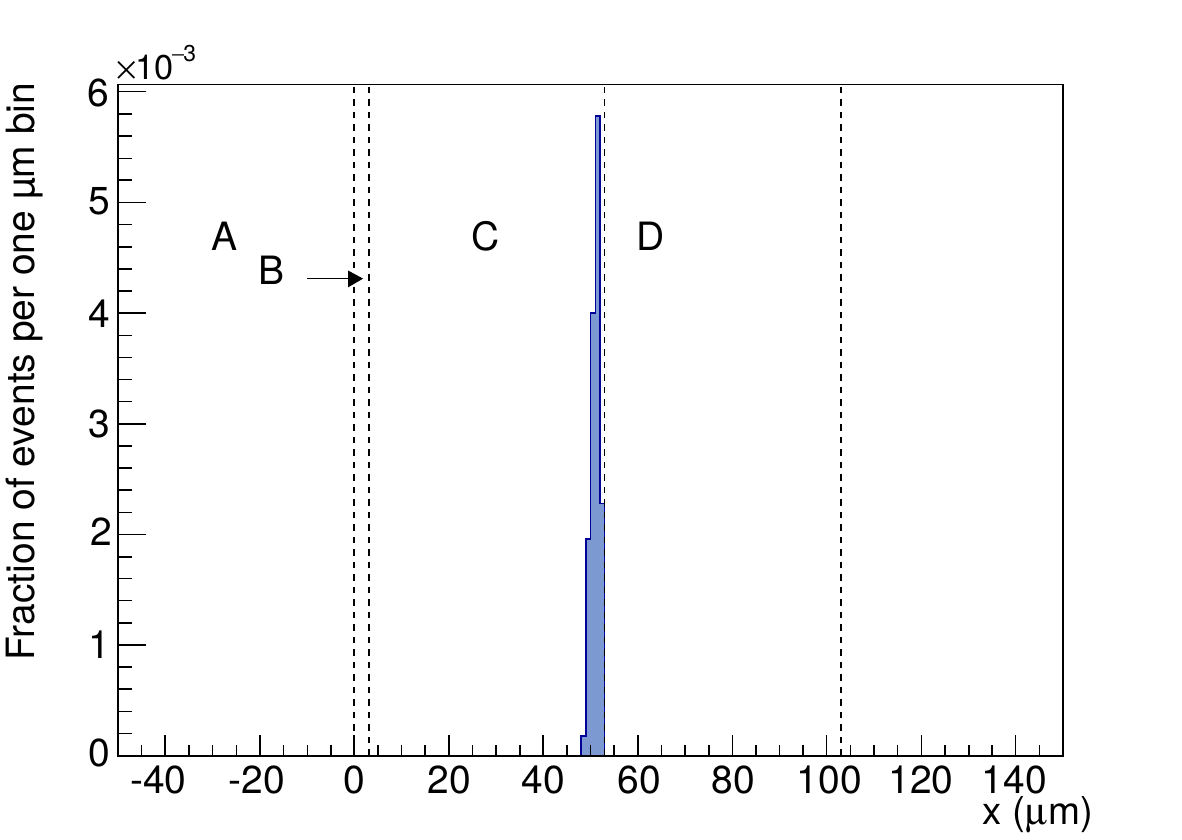}
\includegraphics[width=\figwidth]{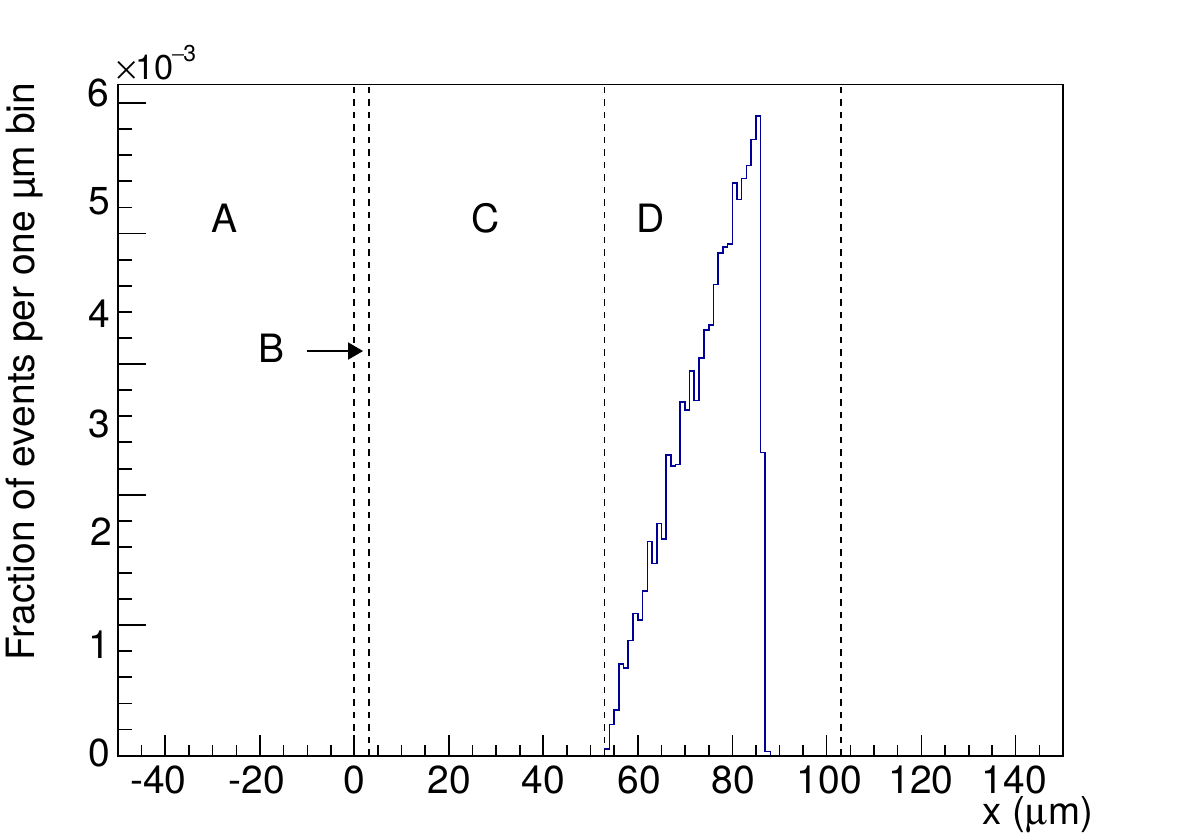}
\caption{\label{scint_ydists}Distribution in $x$ (as defined in Fig.~\ref{concept}) of events which populate the low-energy region of interest (left) for events generated in the scintillator and (right) for events generated in the acrylic.}
\end{figure*}

\begin{figure}
\includegraphics[width=\figwidth]{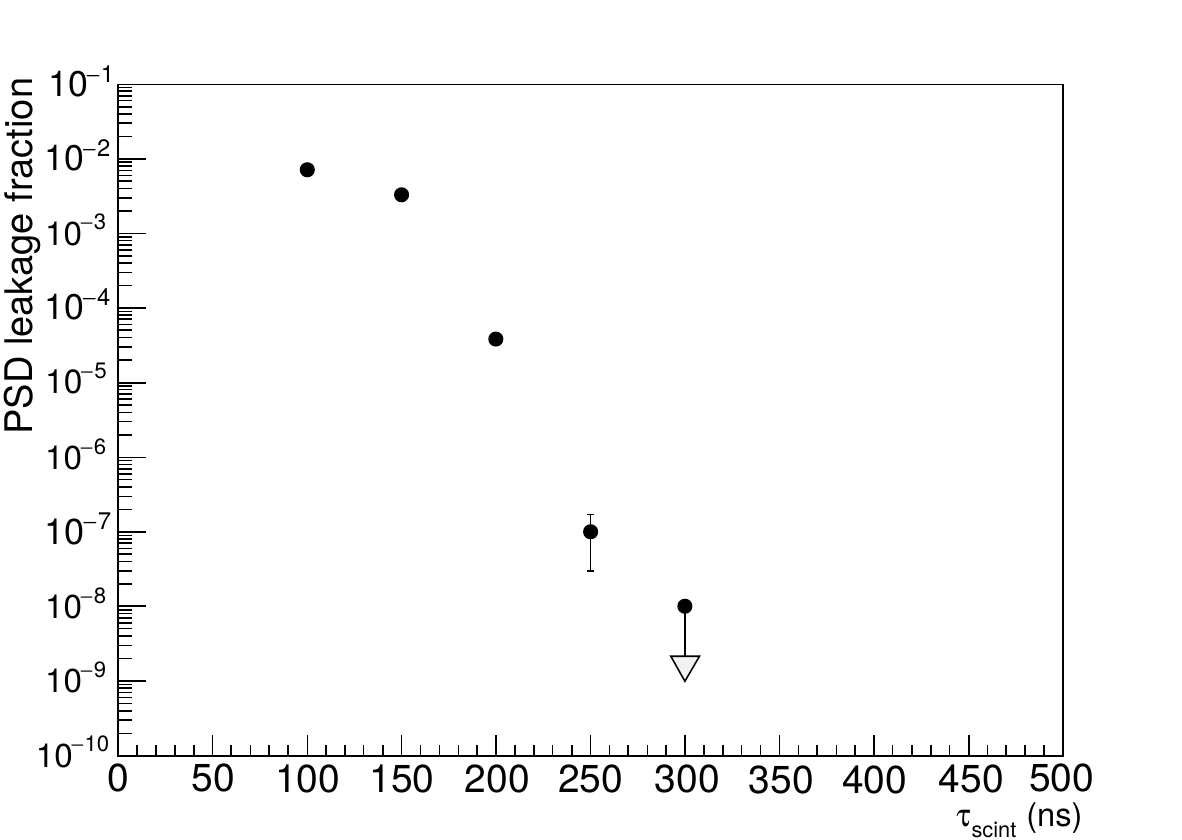}
\caption{\label{rejvstau}Surface background rejection versus the scintillator time constant.  For time constants larger than about 300~$\mu$s, pulse-shape discrimination is greater than $10^{8}$. }
\end{figure}

When referring to energies of nuclear recoils, units of either keV$_{ee}$ (``electron equivalent'') or keV$_{nr}$ are used, with the latter being the full energy of the recoil, and [keV$_{nr}$]=$L_{eff}\cdot$[keV$_{ee}$].
To define a region of interest,  low-energy nuclear recoils with 60~keV kinetic energy (KE) were simulated in the liquid argon.  Figure~\ref{scint_nr} shows the distribution of F$_{prompt}$ versus the total number of detected photons (photoelectrons, PE) and the equivalent observed recoil energy (keV$_{nr}$).    The box in Fig.~\ref{scint_nr} corresponds to a dark matter search window, typical for a liquid-argon-based detector, which accepts 99\% of the low-energy nuclear recoils.
\footnotetext{A recent measurement of NR quenching for $^{206}$Pb~\cite{lowquench} in liquid argon finds a value much lower than expected and from earlier measurements~\cite{hitachi,deap1_radon_paper}.  We use here a nominal value of 0.2 and find that the results of this study are insensitive to this quenching factor.}
Figure~\ref{scint_tpb_surf} shows the simulation result for $^{210}$Po decays generated at the surface of the TPB layer, expected for the case when progeny of $^{222}$Rn are released in the argon and then adsorb to the surface layer.  The left panel shows the case without any scintillator (region C simulated as acrylic), where a significant fraction of the events are found to populate the low-energy recoil region. The right panel shows the simulation result including region C, with a time constant of 500~$\mu$s.   With the scintillator layer in place, all regions in which $\alpha$-particles can deposit energy are active, and so in all cases the observed energy is significantly above the low-energy region of interest.   Without the scintillating layer, events from this region are particularly problematic, since roughness of the TPB surface, expected for thermally-deposited thin films~\cite{tpb_deposition_paper} leads to a difficult-to-model sharing of energy deposition between liquid argon and TPB~\cite{deap1_radon_paper}, both with different scintillation time constants.  With the scintillating layer in place, events are effectively discriminated against, primarily by ensuring that the observed energy is above the region of interest.  Similarly, events generated throughout the TPB layer (region B) will always deposit all of their energy in an active region and can be discriminated based on their observed energy, as seen in Fig.~\ref{scint_tpb}.
Events generated in regions C or D can lose some of their energy in the inactive region D, and leave a small energy deposited in the scintillator (layer C), and so end up with energies in the region of interest.  In these cases the slow decay time of the scintillator can be used to discriminate these events using PSD. Figure~\ref{scint_scint} shows a simulation for events generated uniformly in the scintillator (region C) and  in the acrylic (region D).   Low-energy events are those where most of the $\alpha$ energy is lost to the acrylic.  Figure~\ref{scint_ydists} shows the distribution along $x$ (as defined in Fig.~\ref{concept}) of these events.  
For events generated in the scintillator, only a small band near the interface between the scintillator and acrylic allows these events.   Events anywhere from the acrylic layer can lead to low energy deposition, either where most of the energy is lost in the acrylic near the scintillator, and a small-angle scatter into the scintillator deposits the appropriate energy, or from particles that lose most of their energy while traveling from a lateral distance that is close to their range.    The level of discrimination for this class of events was evaluated as a function of the scintillator decay time and is shown in Fig.~\ref{rejvstau}.  For decay times greater than approximately 300~ns, the discrimination power is greater than $10^{8}$.  Effective discrimination does require the scintillator to have sufficient light output, which is guaranteed as long as observed energies of events from region B (TPB) are above the region of interest. In this case, the PSD strength for events in regions C and D (scintillator and acrylic) will be determined mostly by the decay time (statistics of prompt and late photon counting), and so will be relatively independent of the yield.  

\section{Discussion}
\label{discussion}
We have shown that a layered configuration including a thin layer of scintillating material can effectively mitigate surface $\alpha$-decays in an argon detector as a background for low-energy nuclear recoils.   This configuration allows significantly greater discrimination than is possible  using only the TPB layer itself, either by making the layer thick enough to increase the total light yield from $\alpha$-particles, or by using the slow scintillation response of TPB~\cite{tpb_psd_princeton,veloce_tpb,pollmann_tpb,derek_thesis}.

There are several options for scintillating materials with the required timing properties.  In general, scintillators are usually selected for {\it{fast}} timing to allow high event rates, but several slow scintillators are known, such as BC-444~\cite{bc444} with 285~ns decay time (for other examples see~\cite{long_decay_times}).  It is known that decay time constants are typically temperature-dependent, and for a majority of scintillators decay times increase as temperature decreases from room temperature.  In addition to pure scintillating layers, it is also known that scintillators can be mixed within a polymethyl methacrylate (PMMA) matrix~\cite{acrylic_scintillators}, and so it may be possible to simply dope a thin outer layer of PMMA and coat it with a TPB film to achieve the configuration studied here.  An example candidate material is CaF$_{2}$, an efficient scintillator for $\alpha$-particles with a time constant of $\approx$ 900~ns and a relatively high light yield.  While the light output of scintillators can be temperature-dependent in a non-trivial and non-monotonic way, for many materials, including CaF$_{2}$~\cite{caf2}, it remains sufficiently high at liquid-argon temperature or even increases with decreasing temperature.  CaF$_{2}$ is also commonly deposited as a thin film in order to produce anti-reflective coatings.

To fully exploit discrimination of surface $\alpha$-decays as described here, we will need to identify a candidate scintillator with an appropriately long time constant, high light yield for $\alpha$'s but sufficiently low yield for scintillation due to electrons, and for the case of a 200~m$^{2}$ surface area detector, radiopurity that limits the total rate of $\alpha$-decays to be low enough to be mitigated with PSD.  We are planning an exhaustive study to identify and test potential scintillating film candidates for these qualities and ultimately evaluate the discrimination power of surface events in a liquid argon detector and the feasibility of achieving a sufficient level of radiopurity.

In summary, we have studied and presented a technique using a combination wavelength shifter and scintillating thin film arrangement that may allow powerful suppression of surface backgrounds in a liquid argon detector, with suppression factors of 10$^{8}$ or greater. The technique allows background suppression without requiring position reconstruction and the use of a restricted sensitive region, thus allowing the use of most of the detector mass as a sensitive region.  The level of suppression allows mitigation of surface backgrounds in a sensitive dark matter search with 200~m$^2$ of surface area, corresponding to several hundred tonnes of argon, useful for a high-mass dark matter particle search with sensitivity to the neutrino floor.  The technique could also be used for background suppression in any liquid argon detector searching for low-energy nuclear recoils, including smaller detectors aimed at observing coherent neutrino-nucleus interactions. It would enable sensitive low-background measurements of neutron fluxes even for the case of compact detectors with small photo-sensitive area and without position reconstruction.
\section*{Acknowledgements}
M.~K. is grateful for support from the Arthur B.~McDonald Canadian Astroparticle Research Institute and from the International Research Agenda Programme AstroCeNT (MAB/2018/7) funded by the Foundation for Polish Science (FNP) from the European Regional Development Fund.
\bibliographystyle{elsart-num}
\bibliography{ar}
\end{document}